\documentclass[twocolumn,showpacs,preprintnumbers,superscriptaddress,amsmath,amssymb,aps,pre]{revtex4}

\usepackage{graphicx}
\usepackage{dcolumn}
\usepackage{bm}
\usepackage{color}
\usepackage{multirow}

\begin{document}


\title{Dynamic structure of stock communities: A comparative study between stock returns and turnover rates}

\author{Li-Ling Su}
 \affiliation{School of Business, East China University of Science and Technology, Shanghai 200237, China} %
\author{Xiong-Fei Jiang}
 \affiliation{College of Information Engineering, Ningbo Dahongying University, Ningbo 315175, China} %
\author{Sai-Ping Li}
 \email{spli@phys.sinica.edu.tw}
 \affiliation{Institute of Physics, Academia Sinica, Taipei 115 Taiwan} %
\author{Li-Xin Zhong}
 \affiliation{School of Finance, Zhejiang University of Finance and Economics, Hangzhou 310018, China} %
\author{Fei Ren}
 \email{fren@ecust.edu.cn}
 \affiliation{School of Business, East China University of Science and Technology, Shanghai 200237, China} %
 \affiliation{School of Science, East China University of Science and Technology, Shanghai 200237, China} %
 \affiliation{Research Center for Econophysics, East China University of Science and Technology, Shanghai 200237, China} %

\date{\today}

\begin{abstract}
The detection of community structure in stock market is of theoretical and practical significance for the study of financial dynamics and portfolio risk estimation. We here study the community structures in Chinese stock markets from the aspects of both price returns and turnover rates, by using a combination of the PMFG and infomap methods based on a distance matrix. We find that a few of the largest communities are composed of certain specific industry or conceptional sectors and the correlation inside a sector is generally larger than the correlation between different sectors. In comparison with returns, the community structure for turnover rates is more complex and the sector effect is relatively weaker. The financial dynamics is further studied by analyzing the community structures over five sub-periods. Sectors like banks, real estate, health care and New Shanghai take turns to compose a few of the largest communities for both returns and turnover rates in different sub-periods. Several specific sectors appear in the communities with different rank orders for the two time series even in the same sub-period. A comparison between the evolution of prices and turnover rates of stocks from these sectors is conducted to better understand their differences. We find that stock prices only had large changes around some important events while turnover rates surged after each of these events relevant to specific sectors, which may offer a possible explanation for the complexity of stock communities for turnover rates.
\end{abstract}

\pacs{89.65.Gh,89.75.-k,89.75.Hc}

\maketitle

The stock market is a typical complex system with different types of interactions between individuals and listed companies. To understand how the returns of different companies are correlated with each other and identify their community structure is of crucial importance for the study of financial dynamics and portfolio risk estimation. For this purpose the study of stock correlations has attracted much interest \cite{Plerou-Gopikrishnan-Rosenow-Amaral-Stanley-1999-PRL,Laloux-Cizean-Bouchaud-Potters-1999-PRL,
Zheng-Podobnik-Feng-Li-2012-SR,Billio-Getmansky-Lo-Pelizzon-2012-JFE,Kritzman-Li-Page-Rigobon-2011-JPM,
Mantegna-1999-EPJB,Garas-Argyrakis-2009-EPL,Cai-Zhou-Zhou-Zhou-2010-IJMPC,Tumminello-Lillo-Mantegna-2010-JEBO,
Kwapien-Drozdz-2012-PR,Ren-Zhou-2014-PLoS,Jiang-Chen-Zheng-2014-SR,Chen-Mai-Li-2014-PA,Yang-Zhu-Li-Chen-Deng-2015-PA}. It is noted that there is a coexistence of random and collective interactions among stocks in financial markets. The majority of eigenvalues of the stock correlation matrix agree well with the predictions of random matrix theory (RMT) \cite{Plerou-Gopikrishnan-Rosenow-Amaral-Stanley-1999-PRL}, while a few large eigenvalues contain information about the co-movements of particular stocks within specific industry sectors or communities \cite{Plerou-Gopikrishnan-Rosenow-Amaral-Guhr-Stanley-2002-PRE,Shen-Zheng-2009a-EPL}. Revealing the interactions between stocks and their variance over time has provided useful information for portfolio optimization and systemic risk estimation \cite{Zheng-Podobnik-Feng-Li-2012-SR,Billio-Getmansky-Lo-Pelizzon-2012-JFE,Kritzman-Li-Page-Rigobon-2011-JPM,Farmer-1999-CSE,Mantegna-Stanley-2000,Bouchaud-Potters-2000}.

Among the various methods used in detecting stock correlations, Planar Maximally Filtered Graph (PMFG), an extension of Minimal Spanning Tree (MST) is proposed as an efficient approach to file out the internal structure between complex data sets \cite{Tumminello-Aste-DiMatteo-Mantegna-2005-PNAS}. It has been used to study the collective behavior of stock prices in the US equity market, and a cluster formation associated with economic sectors, is quantitatively investigated \cite{Tumminello-DiMatteo-Aste-Mantegna-2007-EPJB}. In a similar work conducted by this method, a structure change was found during the 2008-2009 financial crisis \cite{Aste-Shaw-Matteo-2010-NJP}. It has also been used to investigate the correlation among 57 stock market indices around the world, and both fast and slow dynamics are discovered in the correlation-based PMFG graphs \cite{Song-Tumminello-Zhou-Mantegna-2011-PRE}. From the topological structure of PMFG graph, one can observe the connections between stocks and detect its local communities. On the other hand, an information theoretic (infomap) method was proposed to capture the community structure in complex networks \cite{Girvan-Newman-2002-PNAS,Rosvall-Bergstrom-2008-PNAS}. Unlike the traditional way of identifying community structures, this method makes use of the valuable information of weights and direction of the links to decompose the network into modules. It has been used to reveal the community structure in stock networks constructed by PMFG method, in which the link weights are associated with the correlation coefficients \cite{Jiang-Chen-Zheng-2014-SR,Song-Tumminello-Zhou-Mantegna-2011-PRE}. To better extract the local interactions between the business sectors, the stock correlation matrix is decomposed based on RMT, and the background noises and the global price movement are removed.

Previous studies of stock correlations are mainly concentrated on the return series. It is worth to extend the correlation study to trading volumes, which is known to be an important variable reflecting the liquidity of financial markets. It has been proven that there exists a scaling relationship between trading volumes and price returns \cite{Hasbrouck-1991-JF,Gopikrishnan-Plerou-Gabaix-Stanley-2000-PRE,Gabaix-Gopikrishnan-Plerou-Stanley-2003-Nature,
Lillo-Farmer-Mantegna-2003-Nature,Lim-Coggins-2005-QF,Eisler-Kertesz-2006-PRE,Zhou-2012-QF}, and they have universal properties like the fat-tailed distribution and the long-term memory effect \cite{Gopikrishnan-Plerou-Gabaix-Stanley-2000-PRE,Gabaix-Gopikrishnan-Plerou-Stanley-2003-Nature}. The study of the community structures identified by trading volumes is valuable for understanding the interactions among stocks from the aspect of market liquidity, and its comparison with the structure of return series can further help to reveal the relationship between trading volumes and price returns. Much work has been done on analyzing the cross-correlation between price returns and trading volumes by various methods, e.g. Granger causality test \cite{Chen-2012-JBF}, detrended cross-correlation analysis \cite{Kristoufek-2015-PA}, and synchronized cross-correlation \cite{Chen-Qiu-Jiang-Zhong-Wu-2015-PA}. However, the study about the correlation between trading volumes is still rare \cite{Lee-Hwang-Kim-Koh-Kim-2011-PA,Zeng-Lemoy-Alava-2014-JSMTE}. The topological property of the MST graph constructed from buy and sell volumes on the Korean stock market has been studied \cite{Lee-Hwang-Kim-Koh-Kim-2011-PA}. By using the inference techniques, it has been shown recently that the traded volumes can provide the information about market mode and interactions within industry sectors \cite{Zeng-Lemoy-Alava-2014-JSMTE}.

In this paper, we study the dynamic structure of stock communities based on the cross-correlations between stocks on the Chinese stock market, by using a combination of the PMFG and infomap methods. A comprehensive comparison is performed between community structures measured by price returns and turnover rates. To further study the dynamic evolution of community structures, we study the stock communities measured by price returns and turnover rates in five time periods, each having a time span of about two years, and explain their differences according to the specific conditions of the stock markets in that period.

\section{Data and Methods}

\subsection{Data}

The dataset used in our study is retrieved from the Beijing Gildata RESSET Data Technology Co., Ltd, see http://www.resset.cn/. Our database contains the daily closing prices and trading volumes of all A-Share stocks traded on Shanghai Stock Exchange (SHSE), one of the two stock exchanges in mainland China. The A-Share stocks are issued by mainland Chinese companies, and traded in Chinese Yuan. Among the stocks of financial industry, more than half of banks began their IPO on SHSE after 2007. To ensure that our sampling stocks are evenly distributed in major important conventional industries, we select the A-share stocks traded on Shanghai Stock Exchange from October 8, 2007 to March 31, 2015 covering the periods of the subprime mortgage crisis in 2008 and European debt crisis started near the end of 2009. We further select the most liquidity stocks traded on the stock exchange to ensure that the stocks have enough number of trading days to be statistically significant in our studies. The filtering criterion is to exclude those stocks that are continuously suspended from the market for more than 10 days or suspended from the market for more than 30 days in total. This filtering yields the sample dataset of 350 A-Share stocks and 625412 daily records in total.

According to the Global Industry Classification Standard (GICS), the stocks traded on Shanghai Stock Exchange are categorized into nine sectors: energy, materials, industrials, consumer goods, health care, banks, real estate, information technology, and utilities. Banks and real estate are two important sectors in the financial industry, and we make separate analysis here. Table~\ref{TB:Industry:Summary} is the summary statistics of the 9 industry sectors, including the names and codes of sectors, and the number of chosen stocks belonging to each sector. Note that we also include a conceptional sector, called "New Shanghai" (SH), which is found to be one of the fundamental sectors composing stock communities. The New Shanghai stocks refer to those stocks issued by companies dealing with financial affairs and urban constructions in Shanghai.

\begin{table}[htp]
\centering
\caption{
The industry sectors are categorized based on the Global Industry Classification Standard (GICS). The basic information includes the name and code of the industry sector, and the number of chosen stocks from each sector.}
\label{TB:Industry:Summary}
\begin{tabular}{lp{1.5cm}c}
  \hline
  Name & Code & Number of stocks \\
  \hline
    Energy                     & EN      & 11  \\
    Materials                  & MA      & 60  \\
    Industrials                & IN      & 82 \\
    Consumer goods             & CG      & 86 \\
    Health care                & HC      & 30  \\
    Banks                      & BA      & 11  \\
    Real estate                & RE      & 26  \\
    Information technology     & IT      & 27 \\
    Utilities                  & UT      & 17 \\
  \hline
\end{tabular}
\end{table}

The trading volume is known to be an important variable reflecting the liquidity of financial markets, and is a non-stationary series. Owing to the non-stationarity property of the trading volume, we consider
the percentage changes in volume in our empirical analysis, well-known as the turnover rate. The turnover rate is defined as the ratio of the number of shares of a stock traded at time $t$ to the total number of outsanding shares for the stock. Hence, we use turnover rates to study the stock correlations instead of trading volumes, and compare with the correlations obtained from price returns.

\subsection{Correlation coefficient matrix decomposition}

We use the Pearson's correlation coefficient to quantify the cross-correlations between individual stocks. For the time series of two different stocks $X_i(t)$ and $X_j(t)$, e.g. return series or turnover rate series, the Pearson's correlation coefficient is defined as
\begin{equation}
c_{ij}=\frac{\langle [X_i(t)- \langle X_i(t)\rangle] [X_j(t)- \langle X_j(t)\rangle] \rangle}{\sigma_i \sigma_j}, \label{equ_correlation}
\end{equation}
where $\sigma_i$ and $\sigma_j$ are the standard deviations of the two stock series. If there are $N$ sample stocks, we shall have a correlation matrix $C$ with $N\times N$ correlation coefficients as its elements.

Many studies of stock correlations have shown that most of the eigenvalues in the stock correlation matrix agree well with the predictions of RMT \cite{Plerou-Gopikrishnan-Rosenow-Amaral-Stanley-1999-PRL}.  The analytical results of the random matrices are as follows. For the correlation matrix of $N$ random time series of length $L$, the probability density function (PDF) $P(\lambda)$ of the eigenvalues $\lambda$ in the limit $N\rightarrow\infty$ and $L\rightarrow\infty$ is given by
\begin{equation}
P(\lambda)=\frac{Q}{2\pi}\frac{\sqrt{(\lambda_{max}-\lambda)(\lambda-\lambda_{min})}}{\lambda}, \label{equ_RMT}
\end{equation}
where $Q\equiv L/N >1$, and $\lambda$ is within the bounds $\lambda_{min}\leq\lambda\leq\lambda_{max}$. $\lambda_{min}$ and $\lambda_{max}$ are the minimum and maximum eigenvalues of the random correlation matrix, and are given by
\begin{equation}
\lambda_{min,max}=1+\frac{1}{Q}\mp2\sqrt{\frac{1}{Q}}.
\label{equ_RMT_eigenvalue}
\end{equation}

Notice that there also exist a few eigenvalues whose values are much larger than the eigenvalues of a random correlation matrix. The largest eigenvalue indeed quantifies a market-wide influence on all stocks \cite{Plerou-Gopikrishnan-Rosenow-Amaral-Guhr-Stanley-2002-PRE,Ren-Zhou-2014-PLoS}, and some of the other large eigenvalues contain information about the co-movements of particular stocks within specific industry sectors or communities  \cite{Plerou-Gopikrishnan-Rosenow-Amaral-Guhr-Stanley-2002-PRE,Shen-Zheng-2009a-EPL}.

To better extract the interactions between stock sectors, we follow the procedures in \cite{Jiang-Chen-Zheng-2014-SR}, and decompose the correlation matrix into three modes: random mode, market mode and sector mode. The decomposition equation is as follows
\begin{equation}
\begin{aligned}
c_{ij}&=c^{random}_{ij} + c^{market}_{ij} + c^{sector}_{ij} \\ &= \sum_{\lambda_{\alpha} \leq \lambda_{max}}\lambda_{\alpha} u^{\alpha}_i u^{\alpha}_j + \lambda_0 u^0_i u^0_j + \sum_{\lambda_{\alpha} > \lambda_{max}}\lambda_{\alpha} u^{\alpha}_i u^{\alpha}_j. \label{equ_decomposition}
\end{aligned}
\end{equation}
The random mode is measured by the eigenvalues restricted to the analytical result of the random matrices ($\lambda_{\alpha} \leq \lambda_{max}$) and their associated eigenvectors, and it is analogous to the white noise in stochastic processes. The market mode is quantified by the largest eigenvalue $\lambda_0$, which generally corresponds to the influence of all the stocks in the stock market, known as the market-wide influence. The random noise and market-wide influence are removed from the correlation matrix, and the remainder, defined as sector mode, contains information about the industry sectors or communities. We will use the sector mode to construct the stock network and to analyze its community structure.

\subsection{Network construction method}

The PMFG is the simplest way of building graphs embedded on surfaces with a given genus \cite{Tumminello-Aste-DiMatteo-Mantegna-2005-PNAS,Tumminello-DiMatteo-Aste-Mantegna-2007-EPJB,Aste-Shaw-Matteo-2010-NJP}. A general approach to construct this type of graph is as follows:
(i) First sort the correlation coefficients $c_{ij}$ by decreasing order, and we here take the sector mode of the correlation coefficients.
(ii) Pick out the first (largest) element in the order list, and add the edge between nodes $i$ and $j$ to the graph.
(iii) Pick out the next element and add the edge if the resulting graph is embedded on a surface of genus $g$; otherwise skip it.
(iv) Iterate the process by repeating step (iii) until all pairs of $(i,j)$ have been considered.
In general, the larger the genus is, the greater the amount of original information is preserved in the graph. However, this increases the complexity of the graph. The simplest graph is the one associated with $g=0$, which is a triangulation of
a topological sphere. Such a planar graph is called a planar maximally filtered graph
(PMFG). PMFGs not only have the algorithmic advantage that planarity tests are relatively simple to perform, but can also provide a larger amount of information about the internal structure of a stock market \cite{Tumminello-Aste-DiMatteo-Mantegna-2005-PNAS,Aste-Shaw-Matteo-2010-NJP}.

The use of a distance matrix provides an alternative way to construct the PMFG graph, which has also been widely applied in the study of market structures\cite{Tumminello-DiMatteo-Aste-Mantegna-2007-EPJB,Aste-Shaw-Matteo-2010-NJP,Yan-Xie-Wang-2015-IJMPB}. The distance between nodes $i$ and $j$ is computed as $d_{ij} = \sqrt{2(1-c_{ij})}$, where $c_{ij}$ is the sector mode of the correlation coefficients. According to this formula, the minimum distance corresponds to the maximum correlation. In the original PMFG approach, the maximum correlation coefficient is taken as the edge weight. The same graph can also be obtained by minimizing the alternative weight given by the distance. The main difference in the approach is that we rank the distances $d_{ij}$ by ascending order in step (i), and the first element picked out in step (ii) is consequently the smallest. All other steps remain to be the same.

Theoretically the PMFG graph constructed by the correlation coefficients is the same as the one obtained by the measure of distances, which could be demonstrated easily in our study. Therefore, we will only show the results of the graph constructed by the distance matrix. Recently, a new study has used absolute correlation coefficients to construct PMFG graphs \cite{Jiang-Chen-Zheng-2014-SR}. This method detects the interactions between stocks based on the magnitude of their correlations, regardless of the signs of their correlations. By doing so, those stocks with high positive and negative correlations could be connected through edges, and anti-correlated sectors, between which stocks are negatively correlated, are consequently observed in their community structure \cite{Jiang-Chen-Zheng-2014-SR}. We also use absolute correlation coefficients to construct the PMFG graph for comparison with the graph obtained by the distance matrix.

An infomap method is further used to reveal the community structure in the stock network constructed by the PMFG method. Before doing that, an adjacent matrix needs to be computed according to the PMFG graph. The elements corresponding to the pairs of $(i,j)$ which have edges in the PMFG graph are constructed from the correlation coefficients of the sector mode. For other pairs of $(i,j)$ which do not have edges in the PMFG graph, the elements are set to be zero. We use this adjacent matrix to identify stock communities based on the infomap method. This information theoretic method is proposed to reveal community structures in weighted and directed networks. It makes use of the probability flow of random walks on a network as a proxy for information flows in the real system, and searches for a module partition of network nodes so as to minimize the expected description length of the random walk. For more details about this method, the reader is referred to \cite{Rosvall-Bergstrom-2008-PNAS}.

\section{Community structures in whole period of sample data}

\subsection{Distributions of correlation coefficients and eigenvalues}

We here use the time series of returns and turnover rates of 350 sample stocks from October 8, 2007 till March 31, 2015, and compute their correlation matrices separately based on the Pearson's correlation coefficient estimate. To get a general understanding of the statistical properties of the correlation coefficients, we first analyze the distribution of the elements $c_{ij}$ of the correlation matrix. Fig. 1 is a plot of the probability density function (PDF) $P(c_{ij})$ of the correlation coefficients. Fig. 1 (a) shows $P(c_{ij})$ for returns calculated by Eq.~(\ref{equ_correlation}), represented by circles in the figure. We observe that $P(c_{ij})$ is biased towards positive $c_{ij}$, and its center is located at a value around 0.45. Similar behavior is observed in $P(c_{ij})$ for turnover rates, as shown in Fig. 1 (b). The peak of $P(c_{ij})$ for turnover rates is located around 0.25, a value smaller than that for returns. We also observe a small portion of $c_{ij}$ is distributed at negative values, which is not seen in $P(c_{ij})$ for returns. These differences suggest that the interaction between individual stocks reflected in price changes is on average stronger than that in market liquidity.

\begin{figure}[htb]
\centering
\includegraphics[width=8cm]{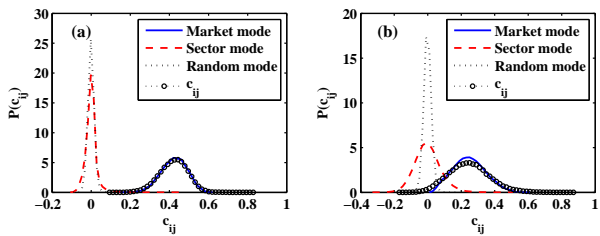}
\caption{\label{Figure-1:CorrDistribution} (Color online) Probability density function (PDF) $P(c_{ij})$ of the correlation coefficients for (a) returns and (b) turnover rates.}
\end{figure}

The correlation coefficient matrix is further decomposed into three modes according to the decomposition Eq.~(\ref{equ_decomposition}). The PDFs of the correlation coefficients for market mode, sector mode and random mode are also shown in Fig. 1, depicted by solid line, dashed line and dotted line respectively. For returns, the PDF corresponding to the market mode gives a curve close to the PDF of the original correlation coefficients, which indicates that the movements of individuals stocks are dominated by the so-called market-wide influence. PDFs for both the random and sector modes have peaks near zero, with the latter having fatter tails especially on the positive side of the horizontal axis. For turnover rates, the PDF for market mode is thinner than that for the original correlation coefficients, and the PDF for sector mode shows tails obviously fatter than that for the random mode. All of these remind us that the collective behavior in market liquidity is not as strong as in price changes and the sector mode for turnover rates contains information more complex than for price returns.

In the calculation of correlation coefficient matrix decomposition, we need to compute eigenvalues of correlation matrices. In Fig. 2, the PDFs $P(\lambda)$ of the eigenvalues of the correlation matrices for returns and turnover rates are plotted. Fig. 2 (a) shows $P(\lambda)$ of the correlation coefficient matrix obtained from the empirical return series (solid line) and the random correlation matrix (dashed line).  The eigenvalues of the random correlation matrix are calculated from Eq.~(\ref{equ_RMT}). The largest eigenvalue $\lambda_0$ used to calculate the market mode $c^{market}_{ij}$ is 149.709. $\lambda_{max}$ of the random correlation matrix is 2.07, and there are eight eigenvalues significantly larger than $\lambda_{max}$.  These large eigenvalues are used to calculate the sector mode $c^{sector}_{ij}$. The other eigenvalues $\lambda \leq \lambda_{max}$ are used to calculate the random mode $c^{random}_{ij}$. The results of turnover rates are shown in Fig. 2 (b). The largest eigenvalue $\lambda_0$ is 93.182, somewhat smaller than the largest eigenvalue for returns, and there are 24 eigenvalues significantly larger than $\lambda_{max}$ of the random correlation matrix.

\begin{figure}[htb]
\centering
\includegraphics[width=8cm]{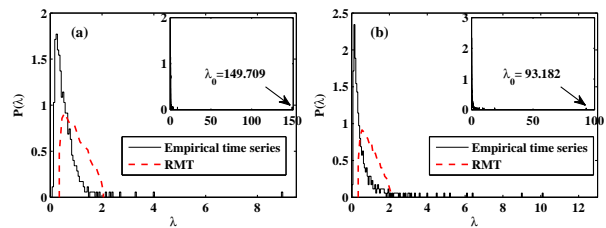}
\caption{\label{Figure-2:EigenvectorDistribution} (Color online) PDF $P(\lambda)$ of the eigenvalues of the correlation matrix for (a) returns and (b) turnover rates.}
\end{figure}

\subsection{Community structures of returns and turnover rates}

We now use the sector mode of the correlation coefficients to compute the distance matrix, and construct the stock network based on the PMFG method. The infomap method is further used to capture the community structure of the PMFG graph. We find that most of the communities in the stock networks comprise particular stocks within specific industry sectors or conceptional sectors which have similar properties. Instead of illustrating the community structures in terms of individual stocks, we draw interactions between sectors for simplicity. Figs. 3 (a) and (b) show the community structures of the PMFG graphs obtained from returns and turnover rates during the period from October 8, 2007 to March 31, 2015. The size of a community represents the sum of the page ranks of all the stocks inside this community, where the page rank of a certain stock reflects its centrality in network. The width of an edge indicates the strength of the interaction between two communities.

\begin{figure}[htb]
\centering
\includegraphics[width=9cm]{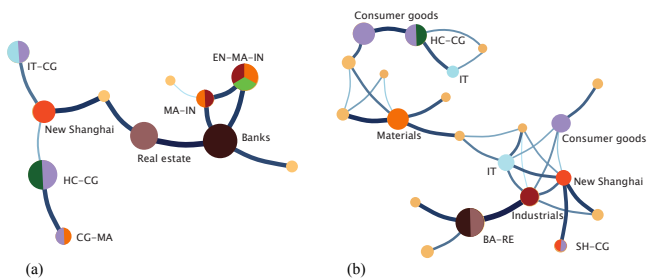}
\caption{\label{Figure-3:community_structure} (Color online) Community structures of the PMFG graphs for (a) returns and (b) turnover rates, using the daily records from October 8, 2007 to March 31, 2015.}
\end{figure}

In the PMFG graphs, there are a few communities composed of multiple sectors, while each of the remaining communities corresponds to one particular sector. In addition, some sectors like consumer goods and materials repeatedly appear in several different communities. The largest community for returns is banks, which is connected to real estate by a strong interaction.  The largest community for turnover rates comprises two sectors, namely banks and real estate. This means that the financial industry which includes banks and real estate play an important role in the Chinese stock market from both aspects of price change and market liquidity.

To better understand the communities extracted from the PMFG graphs, we list the eight largest communities of the PMFG graphs obtained from the correlation matrices of price returns and turnover rates in Table~\ref{TB:community-1}. $N_{stock}$ denotes the number of stocks in each community, and the total number of stocks in the eight communities accounts for a large portion of 350 sample stocks. In Table~\ref{TB:community-1}, the codes of sectors composing these communities are listed, and the number in the bracket after the sector code represents the number of stocks belonging to this sector. The banks sector and consumer goods sector appear in the largest and 2-nd largest communities respectively for both returns and turnover rates. Other sectors, like real estate, materials, IT and New Shanghai, appear in these largest communities for both returns and turnover rates but with different rank orders, showing different positions in the networks for the two time series.

\begin{table}[htp]
\centering
\caption{Eight largest communities of the PMFG graphs obtained from the correlation matrices of price returns and turnover rates using the daily records from October 8, 2007 to March 31, 2015. The stock communities are extracted from the PMFG graphs by the infomap method. The interacting sectors identified in the top 50\% stocks with the highest page ranks and the total stocks of the eight largest communities, the number of stocks $N_{stock}$, the sum of the mean correlation coefficient $\sum_{i}\bar{c}_{i}$ and the sum of the page rank $\sum_{i}PR_{i}$ in each community are listed. The page rank $PR_{i}$ could be regarded as a measure of centrality of stock $i$, and the mean correlation coefficient $\bar{c}_{i}$ is taken average over the correlation coefficients between stock $i$ and its neighboring stocks. The number in the bracket after each sector code denotes the number of stocks from this sector.} \label{TB:community-1}
\resizebox{9cm}{!}{ %
\begin{tabular}{cllccc}
  \hline
  \multicolumn{6}{c}{Return}\\
  \cline{1-6}
  Community & Top 50\% & Total & $N_{stock}$ & $\sum_{i}\bar{c}_{i}$ & $\sum_{i}PR_{i}$\\%
  \hline
    1 & BA(11)          & BA(11)                   & 25 & 0.099 & 0.206 \\
    2 & HC(24) \& CG(7) & HC(28) \& CG(25)         & 61 & 0.080 & 0.165 \\
    3 & EN(8) \& MA(15) & EN(10), MA(23) \& IN(10) & 46 & 0.072 & 0.149 \\
    4 & RE(15)          & RE(19)                   & 35 & 0.068 & 0.140 \\
    5 & IT(18)          & IT(25) \& CG(20)         & 62 & 0.051 & 0.106 \\
    6 & SH(19)          & SH(27)                   & 38 & 0.047 & 0.097 \\
    7 & MA(6)           & MA(9) \& IN(8)           & 27 & 0.027 & 0.055 \\
    8 & CG(9)           & CG(14) \& MA(10)         & 36 & 0.019 & 0.039 \\
  \hline
  \multicolumn{6}{c}{Turnover rate}\\
  \cline{1-6}
  Community & Top 50\% & Total & $N_{stock}$ & $\sum_{i}\bar{c}_{i}$ & $\sum_{i}PR_{i}$\\%
  \hline
    1 & BA(10)          & BA(11) \& RE(6)          & 36 & 0.165 & 0.137\\
    2 & CG(8)           & CG(15)                   & 36 & 0.119 & 0.098\\
    3 & HC(10) \& CG(6) & HC(13) \& CG(11)         & 36 & 0.114 & 0.094\\
    4 & IN(3)           & IN(7)                    & 23 & 0.091 & 0.075\\
    5 & MA(6)           & MA(8)                    & 23 & 0.087 & 0.072\\
    6 & CG(5)           & CG(10)                   & 20 & 0.077 & 0.063\\
    7 & IT(6)           & IT(9)                    & 18 & 0.070 & 0.058\\
    8 & SH(7)           & SH(10)                   & 15 & 0.069 & 0.057\\
  \hline
\end{tabular} }%
\end{table}

One would argue that the sector effect is relatively weak, since the stocks belonging to the identified sectors do not occupy a major proportion of stocks in their corresponding communities.  This is more evident in the case of turnover rates. We further identify the sectors in the top 50\% stocks with the highest page ranks inside the communities, and list the results in Table~\ref{TB:community-1}. We find that the sector effect is more prominent for those stocks at the center of the corresponding communities, which have higher page ranks. Furthermore, the sectors identified in the top 50\% stocks confirm all or part of sectors identified in the total stocks of each community.

The sum of the page rank $\sum_{i}PR_{i}$ in each community is also shown in Table~\ref{TB:community-1}. Since the size of community $i$ is determined by $\sum_{i}PR_{i}$, its value decreases for communities in the order from largest to smallest. The page rank $PR_{i}$ of stock $i$ reflects its centrality in network, which is associated with its interaction strength with other stocks. We assume that the page rank of a stock is proportional to the average value of its correlation coefficients with other stocks. To verify this, we calculate the mean correlation coefficient $\bar{c}_{i}$, which is the average of the correlation coefficients between stock $i$ and its neighboring stocks, and take the sum of the mean correlation coefficient $\sum_{i}\bar{c}_{i}$ over the stocks inside the community. We find that $\sum_{i}\bar{c}_{i}$ decreases as the community size decreases, in agreement with the result of $\sum_{i}PR_{i}$. Therefore, the largest community could be regarded as a cluster of stocks with strongest interactions in total.

To quantitatively understand the interactions between the sectors composing the communities, we also study the correlations inside and between industry sectors in the market. The average correlation inside a sector is defined as
\begin{equation}
\overline{c}^{in}_{ij}= \frac{1}{n_e} c^{in}_{ij} =\frac{1}{n_e} \sum_p \sum_{i,j\in p, i\neq j} c_{ij}, \label{equ_correlation_in}
\end{equation}
where $c_{ij}$ is the correlation coefficient between the stocks from the same sector $p$, $c^{in}_{ij}$ is the sum of the correlation coefficients of edges inside all the sectors, and the average $\overline{c}^{in}_{ij}$ is taken over the total number of edges $n_e$ inside all the sectors. We also measure the average correlation between sectors by
\begin{equation}
\overline{c}^{be}_{ij}=\frac{1}{n'_e} c^{be}_{ij} =\frac{1}{n'_e} \sum_{p\neq q} \sum_{i\in p, j\in q} c_{ij}, \label{equ_correlation_in}
\end{equation}
where $c_{ij}$ is the correlation coefficient between the stocks from sectors $p$ and $q$, $c^{be}_{ij}$ is the sum of the correlation coefficients of edges between different sectors, and the average $\overline{c}^{be}_{ij}$ is taken over the total number of edges $n'_e$ between all the sectors.

In Table~\ref{TB:Correlation:Sector}, the average and sum of the correlation coefficients inside sectors $\overline{c}^{in}_{ij}$ and $c^{in}_{ij}$, and the average and sum of the correlation coefficients between sectors $\overline{c}^{be}_{ij}$ and $c^{be}_{ij}$ are presented. The results are listed for return and turnover rate series by using two methods: method I uses the distance matrix to construct PMFG Graph, that is also the method used in our study; method II uses the absolute value of correlation matrix to construct PMFG Graph, and we present the results using method II in comparison with the results in \cite{Jiang-Chen-Zheng-2014-SR}.

\begin{table}[htp]
\centering
\caption{Average correlation inside and between sectors, denoted by $\overline{c}^{in}_{ij}$ and $\overline{c}^{be}_{ij}$, and the sum of the correlation within and between sectors, denoted by $c^{in}_{ij}$ and $c^{be}_{ij}$. The results are listed for return and turnover rate series by using two different methods: method I makes use of the distance matrix to construct PMFG graph, and method II makes use of the absolute value of the correlation matrix to construct PMFG graph.} \label{TB:Correlation:Sector}
\resizebox{9cm}{!}{ %
\begin{tabular}{llrrrrcrrrr}
  \hline
  & & \multicolumn{4}{c}{$c^{sector}_{ij}$} && \multicolumn{4}{c}{$c_{ij}$}\\
  \cline{3-6}  \cline{8-11}
  & & $\overline{c}^{in}_{ij}$ & $\overline{c}^{be}_{ij}$ & $c^{in}_{ij}$ & $c^{be}_{ij}$  &&  $\overline{c}^{in}_{ij}$ & $\overline{c}^{be}_{ij}$ & $c^{in}_{ij}$ &$c^{be}_{ij}$\\
  \hline
  \multirow{2}{*}{Return} & Method \ I & 0.112 & 0.069 & 48.146 & 18.930 && 0.509 & 0.463 & 218.268 & 127.407 \\
  &                         Method II  & 0.128 & 0.014 & 44.021 & 5.545  && 0.518 & 0.402 & 177.775 & 164.283 \\
  \multirow{2}{1.5cm}{Turnover rate} & Method \ I & 0.222 & 0.202 & 49.378 & 22.573 && 0.484 & 0.467 & 107.547 & 52.296 \\
  &                                Method II  & 0.212 & 0.112 & 40.948 & 21.104 && 0.443 & 0.327 & 85.494 & 61.536 \\
  \hline
\end{tabular} }%
\end{table}

The left panel of Table~\ref{TB:Correlation:Sector} shows the average correlation $\overline{c}^{in}_{ij}$ and $\overline{c}^{be}_{ij}$ calculated from the correlation coefficients of the sector mode $c_{ij}^{sector}$. For returns, $\overline{c}^{in}_{ij}$ is significantly larger than $\overline{c}^{be}_{ij}$ by both methods, which suggests that the correlation between each pair of stocks inside a sector is stronger than the correlation between stocks in different sectors on average. Though this result is similar to that revealed in \cite{Jiang-Chen-Zheng-2014-SR}, we do not observe the anti-correlation between the industry sectors indicated by the negative value of $\overline{c}^{be}_{ij}$. The anti-correlation between sectors is observed only in a few of the largest communities, for instance BA and CG composing the largest community by using method II.  Though the average correlation between sectors is taken over different communities, $\overline{c}^{be}_{ij}$ is a positive value close to zero.  This is because the correlations between sectors in many other communities have small positive values. In fact, $\overline{c}_{ij}^{be}$ shows negative values very close to zero in \cite{Jiang-Chen-Zheng-2014-SR}. The sums of correlations inside and between sectors are also listed in Table~\ref{TB:Correlation:Sector}.  One could see that $c^{in}_{ij}$ is significantly larger than $c^{be}_{ij}$. This suggests that the total correlation between stocks inside sectors is stronger than the correlation between stocks in different sectors.

Similar results are obtained for the turnover rates. $\overline{c}^{in}_{ij}$ and $c^{in}_{ij}$ are respectively larger than $\overline{c}^{be}_{ij}$ and $c^{be}_{ij}$ by method II. The results obtained by method I show that while $\overline{c}^{in}_{ij}$ is only slightly larger than $\overline{c}^{be}_{ij}$, $c^{in}_{ij}$ is much larger than $c^{be}_{ij}$. This result seems reasonable and is consistent with the fact that the communities are ranked by the sum of page ranks associated with the mean correlation coefficients of all the stocks in the community.

The right panel of Table~\ref{TB:Correlation:Sector} shows the average correlation $\overline{c}^{in}_{ij}$ and $\overline{c}^{be}_{ij}$, and the sum of correlation $c^{in}_{ij}$ and $c^{be}_{ij}$ calculated by the original correlation coefficients $c_{ij}$. For both returns and turnover rates, $\overline{c}^{in}_{ij}$ is slightly larger than $\overline{c}^{be}_{ij}$ analyzed by both methods. $c^{in}_{ij}$ is much larger than $c^{be}_{ij}$ by using method I, but is slightly larger than $c^{be}_{ij}$ by using method II. In general, the average and total correlations inside the sectors are similar to those correlations between sectors if the original correlation coefficients are used in the calculation. This is quite different from the results calculated by the sector mode of the correlation coefficients, in which the correlation inside the sectors is obviously larger than the correlation between sectors. We may therefore infer that it is better to use the sector mode to capture the community structure in which case the stocks identified in the community have stronger interactions.

We further make a comparison between the interactions of sectors using methods I and II.  The results obtained by the two methods are qualitatively similar with slight quantitative differences. The correlation inside a sector is generally larger than the correlation between different sectors revealed by both methods, though it is more evident in the observation of average correlation by method II. In our study, we consider the interactions between stocks by taking into account the magnitudes and signs of their correlation coefficients, and use method I to construct the PMFG graph.

\section{Evolutions of stock communities and comparison between returns and turnover rates}

\subsection{Partition of sub-periods}

In the following, we will focus on the evolutions of community structures in stock networks. To better understand the variance of community structures over time, we partition the whole period of our sample data into five sub-periods: from October 8, 2007 to March 31, 2009, from April 1, 2009 to September 30, 2010, from October 8, 2010 to March 30, 2012, from April 5, 2012 to September 30, 2013, and from October 8, 2013 to March 31, 2015. The stock communities are then extracted from the PMFG graph in each sub-period. The length of each sub-period is one and a half year, about 360 trading days on average which is larger than the number of our sampling stocks. In so doing, we make sure that the stocks have enough number of trading days to be statistically significant in each sub-period. On the other hand, the length of a sub-period should not be too long since having enough number of sub-periods is beneficial to the investigation of the variance of community structures.

Fig. 4 gives a plot of the Shanghai Stock Exchange Composite Index (SSEC) as a function of time $t$. The five sub-periods separated by dot-dashed lines are also shown. One can see that in the first sub-period, from October 8, 2007 to March 31, 2009, SSEC had a sharp decline caused by the subprime mortgage crisis in 2008. In the second and third sub-periods, from April 1, 2009 to September 30, 2010 and from October 8, 2010 to March 30, 2012, the SSEC was dominated by declines hit by the European debt crisis which lasted for a long period of time.  It did have a temporal rise near the end of 2008 stimulated by the 4 trillion Chinese Yuan stimulus package announced by the State Council of China. The SSEC displayed a rapid rise due to the ample liquidity in the last sub-period from October 8, 2013 to March 31, 2015. The pattern of the SSEC indicates that the stock market is in a particular status in each sub-period, and the partition into five sub-periods is logical.

\begin{figure}[htb]
\centering
\includegraphics[width=8cm]{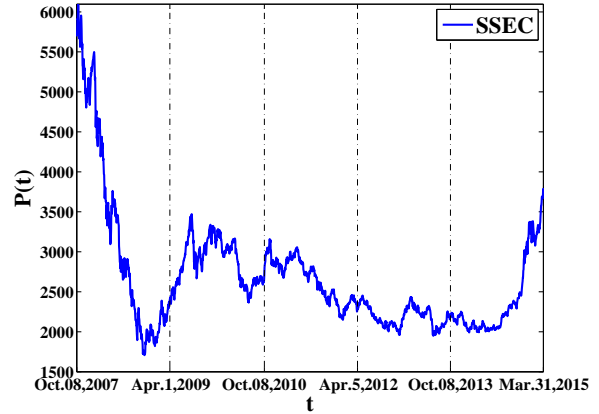}
\caption{\label{Figure-4:price_evolution} (Color online) Evolution of the Shanghai Stock Exchange Composite Index (SSEC).}
\end{figure}

\begin{table}[htp]
\centering
\caption{The largest and smallest eigenvalues of the correlation coefficient matrix, denoted by $\lambda_{0}$ and $\lambda_{349}$, are obtained from the empirical return and turnover rate series in the five sub-periods: from October 8, 2007 to March 31, 2009, from April 1, 2009 to September 30, 2010, from October 8, 2010 to March 30, 2012, from April 5, 2012 to September 30, 2013, and from October 8, 2013 to March 31, 2015. $\lambda_{max}$ and $\lambda_{min}$ are the largest and smallest eigenvalues of the random correlation matrices computed by Eq.~(\ref{equ_RMT_eigenvalue}), and $M$ is the number of the eigenvalues significantly larger than $\lambda_{max}$ of the random correlation matrices.} \label{TB:Eigenvalue:Period}
\resizebox{9cm}{!}{ %
\begin{tabular}{lrrrrr}
  \hline
  \multicolumn{6}{c}{Return}\\
  \hline
  Sub-period & $\lambda_{0}$ & $\lambda_{349}$ & $\lambda_{max}^{ran}$ & $\lambda_{min}^{ran}$ & M\\
  \hline
  October 8, 2007 - March 31, 2009  & 198.413 & 0.001E-1 & 3.933 & 0.003E-1 & 3 \\
  April 1, 2009 - September 30, 2010& 145.909 & 0.002E-1 & 3.907 & 0.005E-1 & 4 \\
  October 8, 2010 - March 30, 2012  & 145.601 & 0.001E-1 & 3.933 & 0.003E-1 & 4 \\
  April 5, 2012 - September 30, 2013& 123.681 & 0.002E-1 & 3.939 & 0.002E-1 & 6 \\
  October 8, 2013 - March 31, 2015  &  97.228 & 0.001E-1 & 3.933 & 0.003E-1 & 6 \\
  \hline
  \multicolumn{6}{c}{Turnover rate}\\
  \hline
  Sub-period & $\lambda_{0}$ & $\lambda_{349}$ & $\lambda_{max}$ & $\lambda_{min}$ & $M$\\
  \hline
  October 8, 2007 - March 31, 2009  & 126.393 & 0.001E-1 & 3.928 & 0.003E-1 & 13 \\
  April 1, 2009 - September 30, 2010&  85.090 & 0.002E-1 & 3.902 & 0.006E-1 & 15 \\
  October 8, 2010 - March 30, 2012  & 107.878 & 0.001E-1 & 3.928 & 0.003E-1 & 15 \\
  April 5, 2012 - September 30, 2013&  61.220 & 0.001E-1 & 3.933 & 0.003E-1 & 16 \\
  October 8, 2013 - March 31, 2015  & 113.466 & 0.001E-1 & 3.928 & 0.003E-1 & 14 \\
  \hline
\end{tabular} }%
\end{table}

Before we study the evolutions of community structures, we would need to decompose the correlation coefficient matrix and compute the part related to the sector mode in each sub-period. Table~\ref{TB:Eigenvalue:Period} provides the basic information about the eigenvalues of the empirical correlation matrices and the random correlation matrices for returns and turnover rates in the five sub-periods. $\lambda_{0}$ and $\lambda_{349}$ are the largest and smallest eigenvalues of the correlation coefficient matrices obtained from the empirical return and turnover rate series. For both return and turnover rate series, $\lambda_{0}$ is largest in the first sub-period, which indicates that the market-wide influence on price changes and market liquidity is strongest during the financial crisis. However, $\lambda_{0}$ does not simultaneously show large values for both time series in any of the other sub-periods, which reflects their differences that exist in collective behaviors. $\lambda_{349}$ has small values in all the sub-periods for both returns and turnover rates. $\lambda_{max}$ and $\lambda_{min}$ are the largest and smallest eigenvalues of the random correlation matrices computed by Eq.~(\ref{equ_RMT_eigenvalue}). $\lambda_{max}$ is around 3.9, and $\lambda_{min}$ is slightly larger than $\lambda_{349}$ of the empirical correlation matrices. $M$ denotes the number of eigenvalues of the empirical correlation matrix that are significantly larger than $\lambda_{max}$ of the random correlation matrix in each sub-period. $M$ generally decreases as $\lambda_{0}$ increases, which has been offered a cursory explanation in \cite{Ren-Zhou-2014-PLoS}. There are 3-6 and 13-16 eigenvalues used in the calculation of sector mode respectively for returns and turnover rates. This further confirms that the sector mode for turnover rates contains information more complex than for price returns.

\subsection{Community structures of returns and turnover rates in different sub-periods}

In the following, we present the community structures of the PMFG graphs obtained from returns and turnover rates in the five sub-periods. Figs. 5 (a) and (b) are plots of the community structures for returns and turnover rates in the {\bf first sub-period} from October 8, 2007 to March 31, 2009. For returns, the largest community comprises banks and industrials, and is connected to two other large communities composed of energy-materials-industrials and real estate respectively. Compared to returns, the community structure for turnover rates is more complex. Its graph has more communities and edges, and the sector effect is relatively weak, having many communities with multiple sector contents. The banks sector again appears in the largest community, but the real estate is in a small community. Furthermore, they are at the edges of the network, not at the center as observed in the graph of returns.

\begin{figure}[htb]
\centering
\includegraphics[width=9cm]{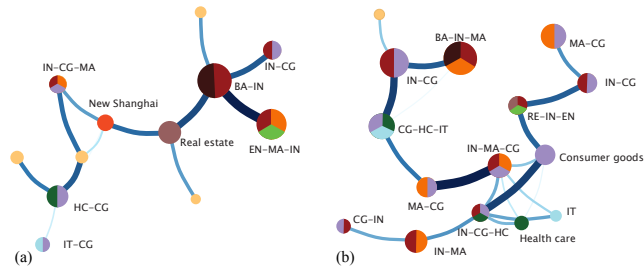}
\caption{\label{Figure-5:community_structure_w1} (Color online) Community structures of the PMFG graphs for (a) returns and (b) turnover rates in the sub-period from October 8, 2007 to March 31, 2009.}
\end{figure}

Table~\ref{TB:community-2} lists the sectors composing the eight largest communities of the PMFG graphs for returns (left) and turnover rates (right) in the five sub-periods. For those communities which do not have clear sector effect, the sector codes are not listed. In the first sub-period, the interacting sectors identified in the top 50\% stocks and the total stocks of the eight largest communities provide similar results. The banks and industrials sectors appear in the largest community for both returns and turnover rates. There also exist obvious differences in the identified sectors for the two time series. For instance, the real estate sector and New Shanghai sector are the 3-rd and 7-th largest communities for returns but they do not appear in the eight largest communities for turnover rates. One can also observe that $\sum_{i}\bar{c}_{i}$ and $\sum_{i}PR_{i}$ decreases as community size gets smaller.

To better illustrate the similarities and differences in the interacting sectors for returns and turnover rates, we show evolutions of prices and turnover rates of the stocks in several specific sectors in the first sub-period in Fig. 6. Banks and real estates are picked as two important sectors, which are associated with the subprime mortgage crisis in 2008. This global financial crisis was triggered by a dramatic rise in default rate on subprime mortgage in US, and caused major adverse effects on real estates and banks all over the world. Figs. 6 (a) and (b) show evolutions of prices and turnover rates for stocks from the banks sector in the largest community for returns and turnover rates respectively, and most of the stocks in the two communities have identical codes. The price and turnover rate are standardized to facilitate comparison between different stocks. Labels A to C in the figure correspond to the following events: A, since September 2008, China reduced the benchmark deposit and lending interest rate five times and deposit reserve ratio four times in that year; B, the State Council of China announced a 4 trillion Chinese Yuan stimulus package on 9 November, 2008; C, in January 2009, the government introduced an industry adjustment and revitalization plan for ten industry sectors. These measures were taken to minimize the impact of the global financial crisis on China. The patterns of the price evolutions for different stocks in Fig. 6 (a) are quite consistent: Before the release of these stimulus measures, the stock prices suffered a disastrous decline affected by the global financial crisis; After imposing these measures, the stock prices reversed their trends. The time evolutions of the turnover rates for different stocks in Fig. 6 (b) exhibit similar behavior, and turnover rates surged after each of these events while the stock prices rose after event C, caused by the cumulative effects of these three events.

\begin{figure}[htb]
\centering
\includegraphics[width=10cm]{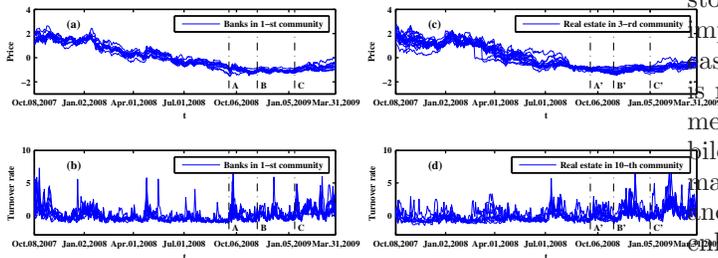}
\caption{\label{Figure-6:w1_BA&RE} (Color online) (a) Evolutions of prices for stocks from the banks sector in the largest community of the graph for returns. (b) Evolutions of turnover rates for stocks from the banks sector in the largest community of the graph for turnover rates. (c) Evolutions of prices for stocks from the real estate sector in the 3-rd largest community of the graph for returns. (d) Evolutions of turnover rates for the stocks from the real estate sector in the 10-th largest community of the graph for turnover rates. All subgraphs are plotted by using data in the sub-period from October 8, 2007 to March 31, 2009.}
\end{figure}

Fig. 6 (c) shows the evolutions of prices for the stocks from the real estate sector in the 3-rd largest community for returns, and Fig. 6 (d) shows the evolutions of turnover rates for the stocks from the real estate sector in the 10-th largest community for turnover rates. Compared to the patterns of the stocks in the largest community, the agreement between the curves for the stocks in either the 3-th or 10-th largest community is relatively weak. Labels A' to C' in the figure correspond to the following events: A', the 4 trillion Chinese Yuan stimulus package was announced on 9 November, 2008; B', from November 2008, the contract tax rate was lowered to 1\% and stamp duty tax and land value-added tax was canceled for personal purchases of housing with a ground floor area of less than 90 square meters, and for the first-time home buyers and upgraders the home loan interest rate was offered a 30\% discount and the minimum down payment was down to 20\%; C', on January 3, 2009, China's four large state-owned commercial banks announced that they might offer 30\% discount on home loan interest rate for the high-quality customers who had applied mortgage loan before the end of October 2008. Although the stock price started to rebound only after event C', turnover rate surged after each of these events. Moreover, the patterns of turnover rates for stocks in the 10-th largest community are clearly different from the patterns of turnover rates for stocks in the largest community, which demonstrates the effectiveness of the infomap method in community detection.

Another specific sector is industrials, which is largely affected by the 4 trillion Chinese Yuan stimulus package. Fig. 7 (a) shows the evolutions of prices for the stocks from the industrials sector in the 2-nd largest community of the graph obtained from returns.  One can see that the patterns of price evolutions for different stocks are in agreement with the listed events. Labels A and B in the figure correspond to the following events: A, the 4 trillion Chinese Yuan stimulus package was announced on 9 November, 2008; B, an industry adjustment and revitalization plan of iron \& steel and automobile industries was issued in January 2009. Similar to the banks and real estate sectors shown above, the stock prices suffered a disastrous decline in the early stage of this sub-period, but reversed their trends after the release of the stimulus package. Most stocks in the 2-nd largest community for returns also appear in the 2-nd and 6-th largest communities for turnover rates, and the evolutions of turnover rates for the stocks in these two communities are also shown.  The patterns of turnover rates for the two communities in Figs. 7 (b) and (c) are very different, which again demonstrates the effectiveness of the infomap method. Turnover rates had a stronger surge after event B than after event A for the stocks in the 2-nd largest community, while stocks in the 6-th largest community exhibit opposite behavior.  Since stocks in the 2-nd largest community are issued by the import and export companies deal in iron \& steel, it is easy to understand that the increase of the turnover rate is more evident after the release of the industry adjustment and revitalization plan of iron \& steel and automobile industries. The stocks in the 6-th largest community mainly belong to the machinery manufacturing industry, and their turnover rates significantly increased due to the enhanced expectations on infrastructure investment after the release of the stimulus package.

\begin{figure}[htb]
\centering
\includegraphics[width=9cm]{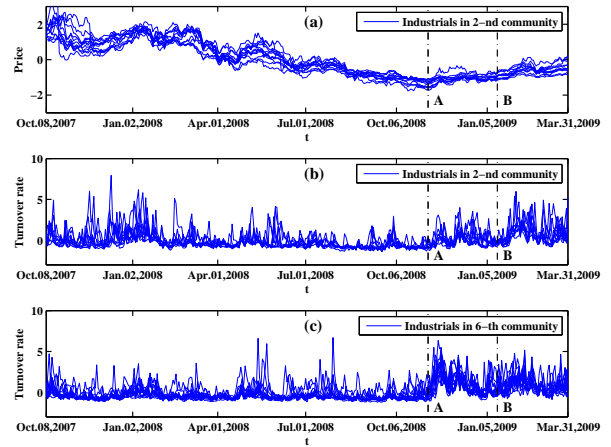}
\caption{\label{Figure-7:w1_IN} (Color online) (a) Evolutions of prices for stocks from the industrials sector in the 2-nd largest community of the graph for returns. (b) Evolutions of turnover rates for stocks from the industrials sector in the 2-nd largest community of the graph for turnover rates. (c) Evolutions of turnover rates for stocks from the industrials sector in the 6-th largest community of the graph for turnover rates. All subgraphs are plotted by using data in the sub-period from October 8, 2007 to March 31, 2009.}
\end{figure}

\begin{figure}[htb]
\centering
\includegraphics[width=9cm]{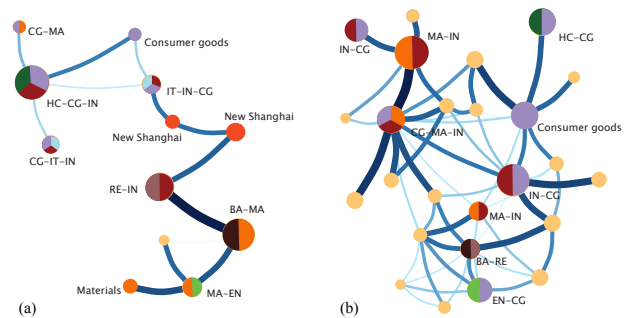}
\caption{\label{Figure-8:community_structure_w2} (Color online) Community structures of the PMFG graphs for (a) returns and (b) turnover rates in the sub-period from April 1, 2009 to September 30, 2010.}
\end{figure}

The community structures for returns and turnover rates in the {\bf second sub-period} from April 1, 2009 to September 30, 2010 are shown in Fig. 8. The community structure for returns in this sub-period keeps its simplicity, similar to the first sub-period: banks and real estate sectors appear in two of the largest communities; and health care, consumer goods and industrials compose the largest community at the edge of the network. The graph for turnover rates is more complex and intensive than in the first sub-period. Sectors like materials, industrials and consumer goods repeatedly appear in the largest communities, and each community is connected to many other communities. Details of the sectors composing the eight largest communities for returns and turnover rates in this sub-period are listed in Table~\ref{TB:community-2}. Health care, banks and real estates appear in the three largest communities for returns respectively, while they appear in relatively smaller communities for turnover rates, i.e., 5-th and 8-th largest communities.

We choose health care as a specific sector, and show the evolutions of prices and turnover rates for the stocks in this sector. Fig. 9 (a) shows the evolutions of prices for the stocks from the health care sector in the largest community for returns, and Fig. 9 (b) shows the evolutions of turnover rates for the stocks from the health care sector in the 5-th largest community for turnover rates. The health care sector had experienced a series of reforms in the second sub-period, and labels A to D in the figure correspond to the following events: A, in April 2009, China released "views of the CPC Central Committee and State Council on deepening the medical and health system", which marked a new round of health-care reform; B, the National Development and Reform Commission (NDRC) released "views on reform of price formation mechanism for drugs and medical services" on November 23, 2009; C, in April 2010, the State Council announced the main work of five reform programs in 2010; D, the NDRC issued drug price control policies in late May 2010. The prices of stocks in health care sector were generally rising in this sub-period due to the expectation of health-care reform, though they experienced a temporal decline around events C and D. The turnover rates remarkably increased around each of these events.

\begin{figure}[htb]
\centering
\includegraphics[width=9cm]{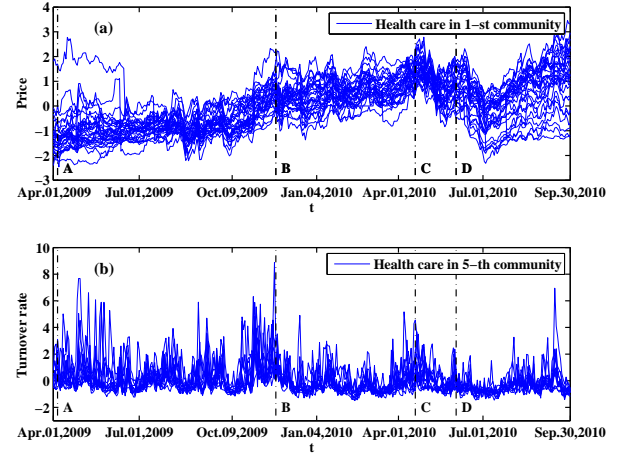}
\caption{\label{Figure-9:w2_HC} (Color online) (a) Evolutions of prices for stocks from the health care sector in the largest community of the graph for returns. (b) Evolutions of turnover rates for stocks from the health care sector in the 5-th largest community of the graph for turnover rates. All subgraphs are plotted by using data in the sub-period from April 1, 2009 to September 30, 2010.}
\end{figure}

\begin{figure}[htb]
\centering
\includegraphics[width=9cm]{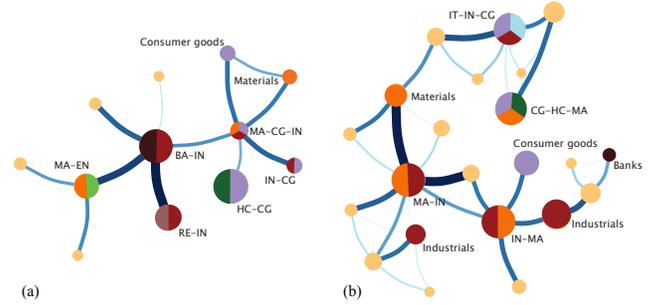}
\caption{\label{Figure-10:community_structure_w3} (Color online) Community structures of the PMFG graphs for (a) returns and (b) turnover rates in the sub-period from October 8, 2010 to March 30, 2012.}
\end{figure}

Fig. 10 shows the community structures for returns and turnover rates in the {\bf third sub-period} from October 8, 2010 to March 30, 2012. The graphs in this sub-period look similar to those in the second sub-period. For returns, health care, banks and real estate respectively appear in the three largest communities. The graph for turnover rates maintains its complexity in which sectors like materials, industrials and consumer goods repeatedly appear in the large communities, but the graph displays a chainlike structure in comparison with the graph in the second sub-period. According to the sectors composing the eight largest communities for returns and turnover rates in this sub-period listed in Table~\ref{TB:community-2}, health care appears in the largest community for returns and the 3-rd largest community for turnover rates. There also exist notable differences between returns and turnover rates: banks and real estate only appear in the eight largest communities for returns, and IT only appears in the eight largest communities for turnover rates.

To better understand their similarities and differences, we choose the health care, banks and IT sectors, and show the evolutions of prices and turnover rates of the stocks from these three sectors. The stock prices were generally falling in this sub-period since the Chinese stock market was deeply affected by the European debt crisis which entered an acute phase from the second half of 2010. However, the turnover rates show entirely different patterns for different sectors. Fig. 11 (a) is a plot of the evolutions of prices for the stocks from the health care sector in the largest community for returns, and the evolutions of turnover rates for the same set of stocks are shown in Fig. 11 (b). Labels A to D in the figure correspond to the following events: A, on November 29, 2010, the NDRC issued an announcement to lower the retail price cap by an average of 19\% among 17 therapeutic drug categories; B, the NDRC set new price ceilings for a list of 162 drugs with an average price cut of 21\% on March 28, 2011; C, China Food and Drug Administration released a notification of implementation of electronic monitoring on essential drugs in 2011 at the end of June 2011; D, China launched the 12th Five-Year Plan for medical device industry on January 18, 2012. The stock prices had sharp declines after events A, B and C, but had a brief turnaround after event D. The turnover rates remarkably increased after each of the four events.

\begin{figure}[htb]
\centering
\includegraphics[width=9cm]{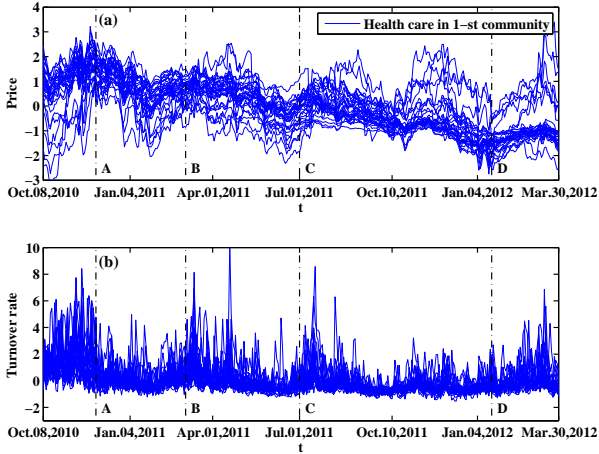}
\caption{\label{Figure-11:w3_HC} (Color online) (a) Evolutions of prices for stocks from the health care sector in the largest community of the graph for returns. (b) Evolutions of turnover rates for the stocks corresponding to those stocks shown in Fig. 11 (a). All subgraphs are plotted by using data in the sub-period from October 8, 2010 to March 30, 2012.}
\end{figure}

Fig. 12 (a) shows the evolutions of prices for stocks from the banks sector in the 2-nd largest community for returns, and the evolutions of their turnover rates are shown in Fig. 12 (b). Labels A to D in the figure correspond to the following events: A, China's Central Bank raised the benchmark deposit and lending interest rate by 0.25\% on October 20, 2010; B, China's Central Bank announced to raise the deposit reserve ratio by 0.5\% points from March 25, 2011; C, Central Huijin Investment Ltd bought shares in four major Chinese state-owned banks on the secondary market from October 10, 2011, aimed at supporting the steady operation and development of major financial institutions and stabilizing their stock prices; D, China's Central Bank cut the deposit reserve ratio by 0.5\% from December 5, 2011. The stock prices had sharp declines after events A and B, and had temporal rises after events C and D. Although the turnover rates do not have clear clustering effect, a surge of turnover rates could also be observed after each of these events.

\begin{figure}[htb]
\centering
\includegraphics[width=9cm]{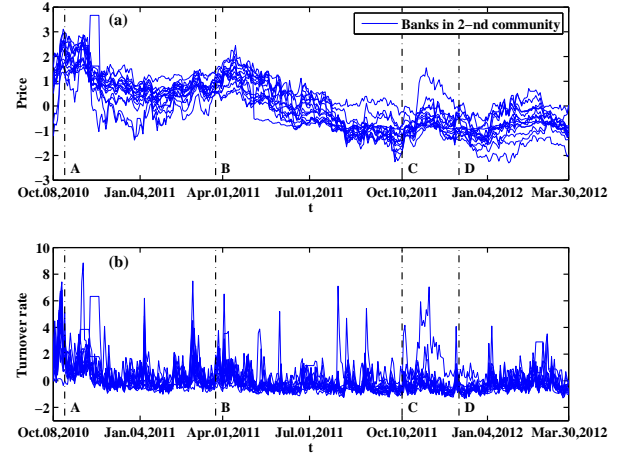}
\caption{\label{Figure-12:w3_BA} (Color online) (a) Evolutions of prices for stocks from the banks sector in the 2-nd largest community of the graph for returns. (b) Evolutions of turnover rates for the same stocks shown in Fig. 12 (a). All subgraphs are plotted by using data in the sub-period from October 8, 2010 to March 30, 2012.}
\end{figure}

\begin{figure}[htb]
\centering
\includegraphics[width=9cm]{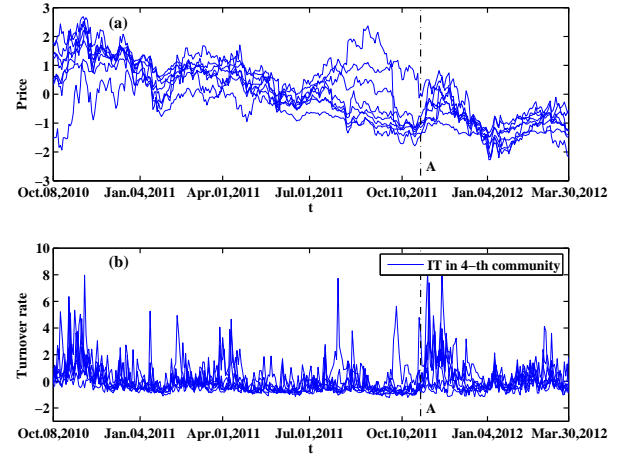}
\caption{\label{Figure-13:w3_IT} (Color online) (a) Evolutions of returns for the stocks corresponding to those stocks shown in Fig. 13 (b). (b) Evolutions of turnover rates for stocks from the IT sector in the 4-th largest community of the graph for turnover rates. All subgraphs are plotted by using data in the sub-period from October 8, 2010 to March 30, 2012.}
\end{figure}

Fig. 13 (b) shows the evolutions of turnover rates for stocks from the IT sector in the 4-th largest community of the graph obtained from turnover rates, and Fig. 13 (a) shows the evolutions of returns for the same stocks shown in Fig. 13 (b). Label A in the figure corresponds to the event that China's State Administration of Taxation and Ministry of Finance announced that enterprises which developed their own software products or redesigned imported software products could have part of their value-added tax refunded on October 13, 2011. The turnover rate remarkably increased immediately after this event.  On the other hand, stock prices had a sharp then transient rise.

In the {\bf fourth sub-period} from April 5, 2012 to September 30, 2013, the community structure for returns retains its simplicity, as shown in Fig. 14 (a). Its largest community comprises health care and consumer goods, and the 2-nd largest community comprises banks and consumer goods, both of which are located at the edges of the network. The large community at the center is the New Shanghai sector. Fig. 14 (b) shows the community structure for turnover rates. The graph is similar to the corresponding graph in the third sub-period, in which sectors like industrials, materials and consumer goods repeatedly appear in large communities. Table~\ref{TB:community-2} also lists the sectors composing the eight largest communities for returns (left) and turnover rates (right) in this sub-period. New Shanghai sector appears in the eight largest communities for both returns and turnover rates, which is not observed in previous sub-periods, and the banks sector is in the 2-nd largest community for returns but not among the eight largest communities for turnover rates. We therefore choose New Shanghai and banks as two specific sectors in this sub-period, and show the evolutions of prices and turnover rates for the stocks from these two sectors.

\begin{figure}[htb]
\centering
\includegraphics[width=9cm]{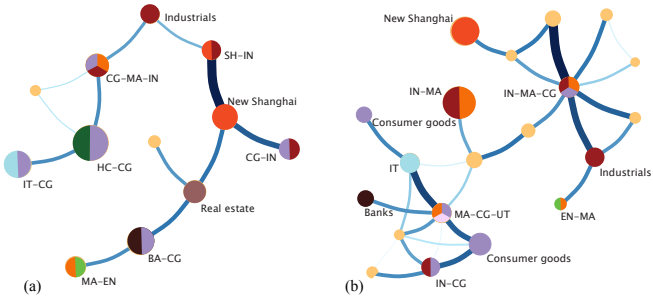}
\caption{\label{Figure-14:community_structure_w4} (Color online) Community structures of the PMFG graphs for (a) returns and (b) turnover rates in the sub-period from April 5, 2012 to September 30, 2013.}
\end{figure}

Fig. 15 (a) shows the evolutions of prices for stocks from the New Shanghai sector in the 4-th largest community for returns, and Fig. 15 (b) shows the evolutions of turnover rates for the stocks from the New Shanghai sector in the 2-nd largest community for turnover rates. Most stocks in these two communities have identical codes. Labels A to C in the figure correspond to the following events: A, the Central Economic Work Conference was held on December 15, 2012, which set the tone for economic priorities for 2013 and beyond; B, on May 22, 2013, the speech given by Ben Bernanke, chairman of the US Federal Reserve, increased the expectation of an interest rate rise, which caused dramatic declines for global stock markets; C, On 22 August, 2013, the State Council of China approved the establishment of Shanghai Pilot Free Trade Zone (SPFTZ). The stock prices continued to decline in the first half of this sub-period, then reversed the trend after event A. Although the stock prices suffered a transient decline after event B, there was a sharp increase after event C. One could also see that the turnover rates remarkably increased during or after each of these events.

\begin{figure}[htb]
\centering
\includegraphics[width=9cm]{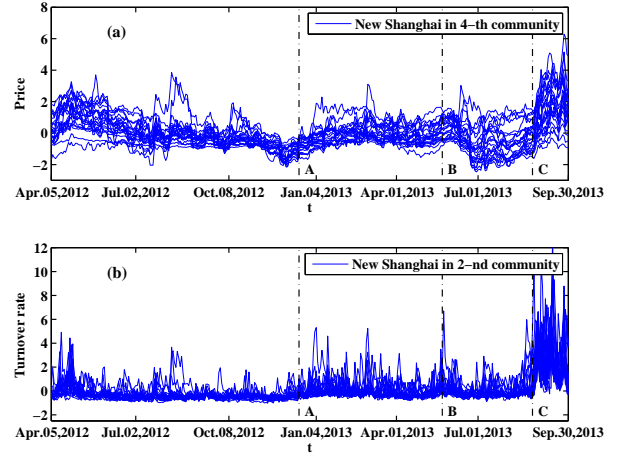}
\caption{\label{Figure-15:w4_SH} (Color online) (a) Evolutions of prices for stocks from the New Shanghai sector in the 4-th largest community of the graph for returns. (b) Evolutions of turnover rates for stocks from the New Shanghai sector in the 2-nd largest community of the graph for turnover rates. All subgraphs are plotted by using data in the sub-period from April 5, 2012 to September 30, 2013.}
\end{figure}

\begin{figure}[htb]
\centering
\includegraphics[width=9cm]{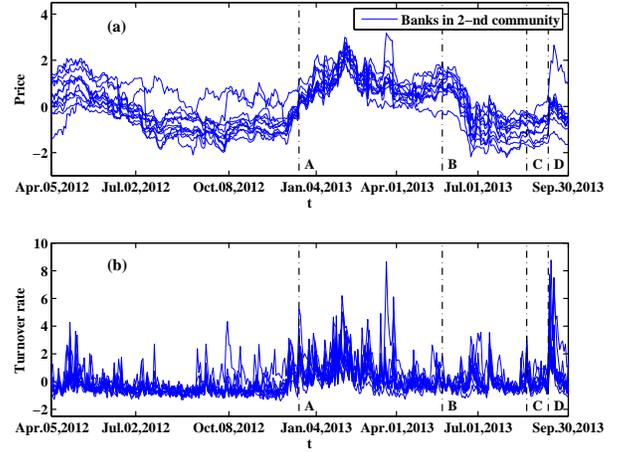}
\caption{\label{Figure-16:w4_BA} (Color online) (a) Evolutions of prices for stocks from the banks sector in the 2-nd largest community of the graph for returns. (b) Evolutions of turnover rates for the same stocks shown in Fig. 16 (a). All subgraphs are plotted by using data in the sub-period from April 5, 2012 to September 30, 2013.}
\end{figure}

Fig. 16 (a) shows the evolutions of prices for stocks from the banks sector in the 2-nd largest community for returns, and the evolutions of their turnover rates are shown in Fig. 16 (b). Labels A and B in the figure correspond to the same events as A and B in Fig. 15: A, the Central Economic Work Conference was held on December 15, 2012; B, the speech given by the chairman of the US Federal Reserve on May 22, 2013. Labels C and D correspond to the following events: C, the SSEC rose more than 5\% in one minute due to the 7 billion Chinese Yuan of fat-finger from Everbright Securities on August 16, 2013; D, Chinese President Xi Jinping raised the initiative of jointly building the Silk Road Economic Belt in September 2013 during his visit to Kazakhstan. Similar to the results of New Shanghai sector, we also observe a rise and a drop in the stock price after events A and B respectively. After an imperceptible and transient decline near event C, the stock prices had a sharp increase after event D while the turnover rates surged during or after each of these events.

\begin{figure}[htb]
\centering
\includegraphics[width=9cm]{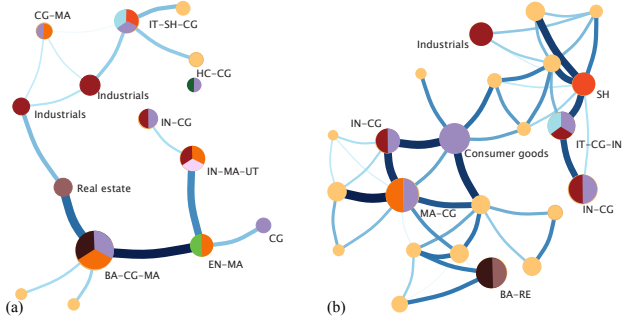}
\caption{\label{Figure-17:community_structure_w5} (Color online) Community structures of the PMFG graphs for (a) returns and (b) turnover rates in the sub-period from October 8, 2013 to March 31, 2015.}
\end{figure}

We now study the community structures for returns and turnover rates in the {\bf fifth sub-period} from October 8, 2013 to March 31, 2015 in Figs. 17 (a) and (b). The graph for returns retains its simplicity, while the graph for turnover rates becomes relatively more complex. We further study the interacting sectors together with the results of sectors composing the eight largest communities listed in Table~\ref{TB:community-2}. Compared to the community structures in previous sub-periods, the main differences lie in the fact that the health care sector disappears from these eight largest communities for both time series and the banks sector reappears in the eight largest communities for turnover rates after a brief absence in the third and fourth sub-periods. We also compare the community structures between returns and turnover rates in the last sub-period.  We observe that the banks, consumer goods and materials sectors are in the top two largest communities for both time series and IT is in the 3-rd and 5-th largest communities for returns and turnover rates respectively. There also exist sectors appearing in communities with different rank orders for two series, e.g., real estate sector appears in the 2-nd largest community for turnover rates and in the 7-th largest community for returns, the New Shanghai sector is in the 3-rd largest community for returns and in the 8-th largest community for turnover rates. The stock prices in these sectors generally have rising trends in this sub-period due to the ample liquidity since the second half of 2014. However, the turnover rates for these sectors show entirely different patterns.

\begin{figure}[htb]
\centering
\includegraphics[width=9cm]{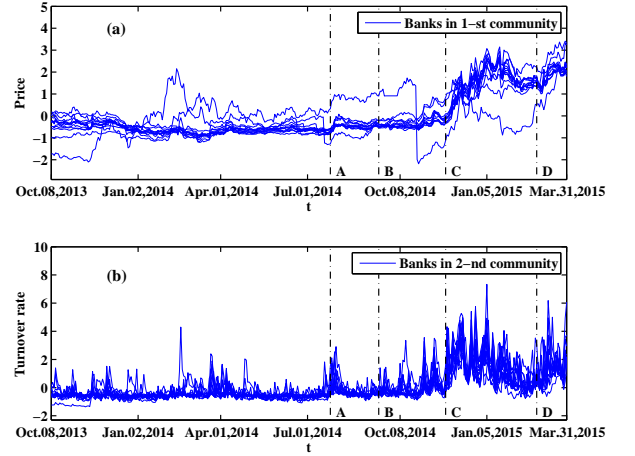}
\caption{\label{Figure-18:w5_BA} (Color online) (a) Evolutions of prices for stocks from the banks sector in the largest community of the graph for returns. (b) Evolutions of turnover rates for stocks from the banks sector in the 2-nd largest community of the graph for turnover rates. All subgraphs are plotted by using data in the sub-period from October 8, 2013 to March 31, 2015.}
\end{figure}

We choose banks, IT and real estate as three sectors for illustration, and show evolutions of their returns and turnover rates as follows. Fig. 18 (a) shows the evolutions of prices for stocks from the banks sector in the largest community for returns, and Fig. 18 (b) shows the evolutions of turnover rates for stocks from the banks sector in the 2-nd largest community for turnover rates. Labels A to D in the figure correspond to the following events: A, China's Securities Regulator announced on July 19, 2014 more plans for the pilot program to connect the Shanghai and Hong Kong stock markets; B, China's Central Bank pumped 769.5 billion Chinese Yuan loans into banks in September and October via a medium-term lending facility; C, on November 22, 2014, China's Central Bank cut the benchmark deposit rate by 0.25\% and the benchmark lending rate by 40\%; D, on March 1, 2015, the benchmark deposit and lending interest rate was reduced by 25\%. Both stock prices and turnover rates show remarkable increases after events C and D. This scenario is evident after event C, since the stock market was also influenced by the launch of the "Stock Connect" link between the Shanghai and Hong Kong stock exchanges on November 17 just before the interest-rate cut on November 22. Although the stock prices were not obviously affected by events A and B, turnover rates surged after these two events.

Fig. 19 (a) shows the evolutions of prices for stocks from the real estate sector in the 7-th largest community for returns, and the evolutions of their turnover rates are shown in Fig. 19 (b). Labels A to D in the figure correspond to the following events: A, China unveiled an urbanization plan for the 2014-2020 period on March 16, 2014, which would bring large demands for urban infrastructure, public service facilities and housing construction; B, China's Central Bank announced on September 29, 2014 that Chinese citizens who wished to buy a second home, would be able to enjoy the same 30\% down payment requirement as first-time home buyers if they had fully repaid their previous mortgage loans; C, on November 22, 2014, China's Central Bank cut the benchmark deposit rate by 0.25\% and the benchmark lending rate by 40\%; D, on March 1, 2015, the benchmark deposit and lending interest rate was reduced by 25\%. Inspired by these four events, the stock prices generally increased in this sub-period. A surge of turnover rates can also be observed after each of these events.

\begin{figure}[htb]
\centering
\includegraphics[width=9cm]{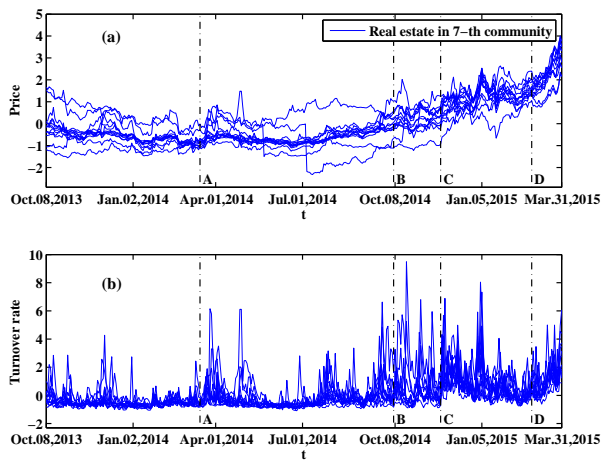}
\caption{\label{Figure-19:w5_RE} (Color online) (a) Evolutions of prices for stocks from the real estate sector in the 7-th largest community of the graph for returns. (b) Evolutions of turnover rates for the same stocks shown in Fig. 19 (a). All subgraphs are plotted by using data in the sub-period from October 8, 2013 to March 31, 2015.}
\end{figure}

Fig. 20 (a) shows the evolutions of prices for stocks from the IT sector in the 3-rd largest community for returns, and Fig. 20 (b) shows the evolutions of turnover rates for stocks from the IT sector in the 5-th largest community for turnover rates. Labels A to D in the figure correspond to the following events: A, on December 4, 2013, China issued 4G network licences to China Mobile, Unicom and Telecom, and China Mobile received license to operate fixed-line broad band services; B, Ministry of Industry and Information Technology (MIIT) and NDRC announced to make broadband China strategy demonstration city list in January 2014; C, the State Council of China published a national guideline for the development and promotion of the integrated circuit industry on June 24, 2014; D, the State Council of China issued a blueprint for China's logistics industry for the 2014-2020 period on September 12, 2014, in which the development of E-commerce Logistics Industry would lead to the development of IT industry. No clear increase in stock prices is observed after event A, while the stock prices remarkably increased after events B, C and D. For turnover rates, the patterns after A-D events are consistent, having a surge after each of these events.

\begin{figure}[htb]
\centering
\includegraphics[width=9cm]{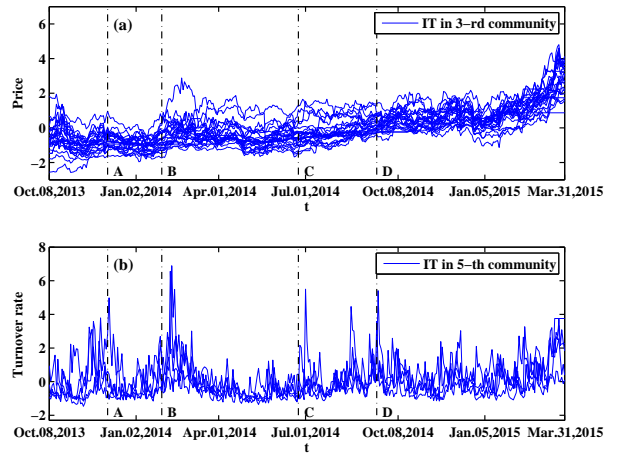}
\caption{\label{Figure-20:w5_IT} (Color online) (a) Evolutions of prices for stocks from the IT sector in the 3-rd largest community of the graph for returns. (b) Evolutions of turnover rates for stocks from the IT sector in the 5-th largest community of the graph for turnover rates. All subgraphs are plotted by using data in the sub-period from October 8, 2013 to March 31, 2015.}
\end{figure}

\begin{table}[htp]
\centering
\caption{Eight largest communities of the PMFG graphs in the five sub-periods: from October 8, 2007 to March 31, 2009, from April 1, 2009 to September 30, 2010, from October 8, 2010 to March 30, 2012, from April 5, 2012 to September 30, 2013, and from October 8, 2013 to March 31, 2015. The stock communities are extracted from the PMFG graphs obtained from the correlation matrices of price returns (left) and turnover rates (right) in each sub-period by using the infomap method. The interacting sectors identified in the top 50\% stocks and the total stocks of the eight largest communities, the number of stocks $N_{stock}$, the sum of the mean correlation coefficient $\sum_{i}\bar{c}_{i}$ and the sum of the page rank $\sum_{i}PR_{i}$ in each community are listed. The number in the bracket after each sector code denotes the number of stocks from this sector. For those communities which do not have clear sector effect, the sector codes are not listed.} \label{TB:community-2}
\resizebox{16cm}{!}{ %
\begin{tabular}{cllccccllccc}
  \hline
  \multicolumn{12}{c}{October 8, 2007 - March 31, 2009}\\
  \hline
  & \multicolumn{5}{c}{Return} && \multicolumn{5}{c}{Turnover rate}\\  %
  \cline{2-6}  \cline{8-12}
  Community & Top 50\% & Total & $N_{stock}$ & $\sum_{i}\bar{c}_{i}$ & $\sum_{i}PR_{i}$ && Top 50\% & Total & $N_{stock}$ & $\sum_{i}\bar{c}_{i}$ & $\sum_{i}PR_{i}$ \\
  \hline
    1 & BA(11)          & BA(11) \& IN(11)    & 33 & 0.088 & 0.197 && BA(5), IN(6) \& MA(5) & BA(7), IN(9) \& MA(13)        & 46 & 0.189 & 0.132\\
    2 & EN(8) \& MA(13) & EN(10), MA(19) \& IN(12)& 45 & 0.075 & 0.166 && IN(6) \& CG(5)   & IN(10) \& CG(13)              & 40 & 0.160 & 0.112\\
    3 & RE(15)          & RE(17)                  & 35 & 0.055 & 0.123 &&                  & CG(10), HC(5) \& IT(6)        & 31 & 0.135 & 0.094\\
    4 & HC(12)          & HC(19) \& CG(9)         & 40 & 0.049 & 0.110 && MA(7) \& CG(7)   & MA(11) \& CG(12)              & 41 & 0.131 & 0.092\\
    5 & CG(8)           & CG(10) \& IN(9)         & 33 & 0.033 & 0.075 && IN(3) \& CG(5)   & IN(5), MA(6) \& CG(7)         & 21 & 0.121 & 0.085\\
    6 & IN(6) \& CG(5)  & IN(12), CG(9) \& MA(8)  & 35 & 0.033 & 0.073 && IN(11)           & IN(14) \& MA(7)               & 31 & 0.105 & 0.073\\
    7 & SH(11)          & SH(19)                  & 22 & 0.029 & 0.065 && CG(6)            & CG(10)                        & 21 & 0.100 & 0.094\\
    8 & IT(13)          & IT(14) \& CG(7)         & 29 & 0.023 & 0.052 &&                  & MA(6) \& CG(4)                & 18 & 0.094 & 0.066\\
  \hline
    \multicolumn{12}{c}{April 1, 2009 - September 30, 2010}\\
  \hline
  & \multicolumn{5}{c}{Return} && \multicolumn{5}{c}{Turnover rate}\\  %
  \cline{2-6}  \cline{8-12}
  Community & Top 50\% & Total & $N_{stock}$ & $\sum_{i}\bar{c}_{i}$ & $\sum_{i}PR_{i}$ && Top 50\% & Total & $N_{stock}$ & $\sum_{i}\bar{c}_{i}$ & $\sum_{i}PR_{i}$ \\
  \hline
    1 & HC(26)          & HC(30), CG(18) \& IN(9) & 66 & 0.135 & 0.212 && MA(4) \& IN(4)   & MA(11) \& IN(6)               & 31 & 0.140 & 0.107\\
    2 & BA(11)          & BA(11) \& MA(8)         & 39 & 0.125 & 0.196 && IN(3) \& CG(4)   & IN(9) \& CG(9)                & 27 & 0.133 & 0.102\\
    3 & RE(14)          & RE(20) \& IN(9)         & 33 & 0.102 & 0.160 &&                  & CG(6), MA(4) \& IN(5)         & 19 & 0.103 & 0.079\\
    4 & MA(5) \& EN (5) & MA(7) \& EN (8)         & 19 & 0.055 & 0.087 && CG(7)            & CG(14)                        & 23 & 0.099 & 0.076\\
    5 & IT(10)          & IT(14), IN(11) \& CG(13)& 45 & 0.052 & 0.082 && HC(11)           & HC(13) \& CG(6)               & 28 & 0.096 & 0.074\\
    6 & SH(18)          & SH(20)                  & 35 & 0.046 & 0.072 &&                  & EN(5) \& CG(7)                & 24 & 0.085 & 0.066\\
    7 & CG(9) \& IT(7)  & CG(15), IT(8) \& IN(14) & 48 & 0.044 & 0.069 &&                  & IN(7) \& CG(6)                & 23 & 0.073 & 0.056\\
    8 & MA(7)           & MA(11)                  & 14 & 0.021 & 0.034 && BA(4) \& RE(2)   & BA(4) \& RE(4)                & 11 & 0.053 & 0.041\\
  \hline
  \multicolumn{12}{c}{October 8, 2010 - March 30, 2012}\\
  \hline
  & \multicolumn{5}{c}{Return} && \multicolumn{5}{c}{Turnover rate}\\  %
  \cline{2-6}  \cline{8-12}
  Community & Top 50\% & Total & $N_{stock}$ & $\sum_{i}\bar{c}_{i}$ & $\sum_{i}PR_{i}$ && Top 50\% & Total & $N_{stock}$ & $\sum_{i}\bar{c}_{i}$ & $\sum_{i}PR_{i}$ \\
  \hline
    1 & HC(26) \& CG(16)& HC(30) \& CG(30)  & 88 & 0.151 & 0.241 &&                        & MA(6) \& IN(7)                & 24 & 0.159 & 0.101\\
    2 & BA(11) \& IN(8) & BA(11) \& IN(18)        & 47 & 0.141 & 0.225 && IN(5) \& MA(3)   & IN(7) \& MA(8)                & 25 & 0.158 & 0.101\\
    3 & RE(17)          & RE(20) \& IN(12)        & 41 & 0.092 & 0.146 && CG(8)            & CG(15), HC(7) \& MA(7)        & 38 & 0.151 & 0.097\\
    4 & MA(13)          & MA(18) \& EN (8)        & 41 & 0.091 & 0.146 && IT(6) \&IN(5)    & IT(8), IN(6) \& CG(9)         & 29 & 0.149 & 0.095\\
    5 & MA(7)           & MA(8), CG(6) \& IN(5)   & 22 & 0.037 & 0.060 && IN(10)           &IN(17)                        & 30 & 0.137 & 0.088\\
    6 & IN(6)           & IN(10) \& CG(6)         & 23 & 0.026 & 0.041 && CG(6)            & CG(8)                         & 19 & 0.098 & 0.062\\
    7 & MA(5)           & MA(10)                  & 16 & 0.020 & 0.033 && MA(7)            & MA(12)                        & 18 & 0.090 & 0.057\\
    8 & CG(7)           & CG(12)                  & 22 & 0.020 & 0.031 &&                  &                               &    &       &      \\
  \hline
  \multicolumn{12}{c}{April 5, 2012 - September 30, 2013}\\
  \hline
  & \multicolumn{5}{c}{Return} && \multicolumn{5}{c}{Turnover rate}\\  %
  \cline{2-6}  \cline{8-12}
  Community & Top 50\% & Total & $N_{stock}$ & $\sum_{i}\bar{c}_{i}$ & $\sum_{i}PR_{i}$ && Top 50\% & Total & $N_{stock}$ & $\sum_{i}\bar{c}_{i}$ & $\sum_{i}PR_{i}$ \\
  \hline
    1 & HC(22)          & HC(29) \& CG(15)        & 54 & 0.137 & 0.195 && IN(8) \& MA(4)   & IN(11) \& MA(11)              & 46 & 0.252 & 0.132\\
    2 & BA(11)          & BA(11) \& CG(7)         & 30 & 0.089 & 0.127 && SH(15)           & SH(25)                        & 40 & 0.227 & 0.119\\
    3 & IT(19) \& CG(7) & IT(23) \& CG(19)        & 63 & 0.083 & 0.118 && CG(6)            & CG(12)                        & 28 & 0.153 & 0.080\\
    4 & SH(15)          & SH(23)                  & 31 & 0.074 & 0.105 &&                  & IN(4), MA(5) \& CG(4)         & 16 & 0.130 & 0.068\\
    5 & CG(8)           & CG(9), MA(8) \& IN(7)   & 30 & 0.067 & 0.095 &&                  & MA(4), CG(4) \& UT(4)         & 18 & 0.124 & 0.065\\
    6 & RE(11)          & RE(16)                  & 25 & 0.063 & 0.090 && IN(4) \& CG(3)   & IN(6) \& CG(5)                & 23 & 0.122 & 0.064\\
    7 & MA(9) \& EN(7)  & MA(17) \& EN(8)         & 33 & 0.054 & 0.077 && IT(7)            & IT(9)                         & 21 & 0.119 & 0.062\\
    8 & CG(7) \& IN(4)  & CG(12) \& IN(7)         & 27 & 0.044 & 0.062 && IN(4)            & IN(9)                         & 20 & 0.105 & 0.055\\
  \hline
    \multicolumn{12}{c}{October 8, 2013 - March 31, 2015}\\
  \hline
& \multicolumn{5}{c}{Return} && \multicolumn{5}{c}{Turnover rate}\\  %
  \cline{2-6}  \cline{8-12}
  Community & Top 50\% & Total & $N_{stock}$ & $\sum_{i}\bar{c}_{i}$ & $\sum_{i}PR_{i}$ && Top 50\% & Total & $N_{stock}$ & $\sum_{i}\bar{c}_{i}$ & $\sum_{i}PR_{i}$ \\
  \hline
    1 & BA(11)          & BA(11), CG(7) \& MA(6)  & 33 & 0.197 & 0.247 && MA(7) \& CG(4)   & MA(13) \& CG(7)               & 29 & 0.139 & 0.099\\
    2 & IN(10) \& MA(6) & IN(17), MA(9) \& UT(9)  & 37 & 0.097 & 0.121 && BA(8) \& RE(3)         & BA(10) \& RE(6)               & 37 & 0.123 & 0.087\\
    3 & IT(12) \& SH(7) & IT(15), SH(11) \& CG(10)& 52 & 0.088 & 0.111 && CG(5)            & CG(9)                         & 23 & 0.122 & 0.087\\
    4 & EN(6) \& MA(3)  & EN(8) \& MA(6)          & 18 & 0.074 & 0.093 && IN(4) \& CG(5)   & IN(8) \& CG(8)         & 30 & 0.116 & 0.082\\
    5 & IN(8)           & IN(18)                  & 42 & 0.066 & 0.083 && IT(4) \& CG(5)                 & IT(5), CG(8) \& IN(6)         & 25 & 0.106 & 0.075\\
    6 & IN(7)           & IN(10) \& CG(9)         & 27 & 0.058 & 0.072 && IN(4)            & IN(6) \& CG(5)                & 16 & 0.084 & 0.060\\
    7 & RE(7)           & RE(13)                  & 19 & 0.051 & 0.064 && IN(5)                 & IN(13)                        & 25 & 0.079 & 0.056\\
    8 & IN(7)           & IN(12)                  & 24 & 0.041 & 0.052 && SH(5)            & SH(6)                         & 13 & 0.077 & 0.054\\
  \hline
\end{tabular} }%
\end{table}

\section{Conclusion}
In this paper, we study the dynamic evolution of the community structures in Chinese stock markets from the aspects of both price returns and turnover rates, using daily data of 350 A-share stocks traded on the Shanghai Stock Exchange from October 8, 2007 to March 31, 2015. The community structure is extracted from the PMFG graph based on the infomap method, and the sector mode decomposed from the correlation matrix is used to better identify stock interactions. The PDF of the sector mode suggests that the sector mode for turnover rates contains information more complex than that for price returns, which is further supported by the result that its eigenvalues are significantly larger than the eigenvalues of the random correlation matrix.

We investigate the community structure by using all daily records, and find that most of the communities in PMFG graphs are composed of definite industry or conceptional sectors. The results of PMFG graph and the listed eight largest communities show that the banks, real estate, health care and New Shanghai sectors compose a few of the largest communities for both return and turnover rate time series, but with different rank orders. These sectors are also observed in the largest communities in the partitioned sub-periods. In addition, the community structure is more complex and the sector effect is relatively weaker for turnover rates than for returns, in agreement with the results of the PDFs of decomposed modes and eigenvalues of the correlation matrix. By analyzing both the average and sum of correlations inside and between the sectors which compose the communities, we find that the correlation inside a sector is generally larger than the correlation between different sectors. This result is robust to different construction methods of PMFG graph, i.e., absolute value of correlation matrix and distance matrix, and the latter is used in our study.

In order to study the dynamic evolution of community structures, we partition the whole period of our sample data into five sub-periods, and study the community structures in each sub-period. Banks sector appears in the largest community for both returns and turnover rates in the first sub-period due to the global financial crisis in 2008, and reappears in the top two largest communities for both time series in the last sub-period accompanied by ample liquidity. Health care sector composes the largest community for returns during the second to fourth sub-periods due to the health-care reform, and also appears in the 3-rd largest community for turnover rates in the third sub-periods. New Shanghai sector appears in the top four largest communities for both time series in the fourth sub-period due to the establishment of SPFTZ. However, there exist several specific sectors which appear in communities with different rank orders for the two time series. We further compare the evolutions of prices and turnover rates of the stocks from these sectors, and offer an interpretation for their differences based on historical events related to these sectors. In our study, we found that turnover rates are more susceptible to external events. Stock prices had large changes only around some important events, while turnover rates surged after each of these events. The difference between the response of returns and turnover rates to exogenous shocks may reflect the complexity of their corresponding network structures. A relationship between the fluctuations of returns and the way that turnover rates respond to exogenous shocks should be an interesting subject worth to study in the future.

\begin{acknowledgments}
The authors would like to thank Wei-Xing Zhou for helpful comments and suggestions. This work was partially supported by the National Natural Science Foundation (Nos. 10905023, 71131007, 71371165 and 11501199), Humanities and Social Sciences Fund sponsored by Ministry of Education of the People¡¯s Republic of China (No. 09YJCZH042), the Shanghai (Follow-up) Rising Star Program Grant 11QH1400800, Ningbo Natural Science Foundation (No. 2015A610160), and the Fundamental Research Funds for the Central Universities.
\end{acknowledgments}

\bibliography{E:/Papers/Auxiliary/Bibliography} 

\end{document}